\documentclass{epl}
\usepackage{amssymb,graphicx}

\newcommand{\R}{{\mathbb R}}

\newcommand{\half}{{\frac{1}{2}}}

\newcommand{\La}{\left\langle}
\newcommand{\Ra}{\right\rangle}
\newcommand{\sLa}{\langle}
\newcommand{\sRa}{\rangle}
\newcommand{\cut}[1]{}

\newcommand{\hxi}{{\hat\xi}}

\title{ Statistical physics of independent component analysis
      }

\author{R. Urbanczik} 
  \institute{
        Institut f\"ur theoretische Physik -
        Universit\"at W\"urzburg,
        Am Hubland,
        D-97074 W\"urzburg,
        Germany       }

     {Order disorder transitions; statistical mechanics of model systems} 

\begin{document}\maketitle

\begin{abstract}
Statistical physics is used to investigate independent component
analysis with polynomial contrast functions. While the replica method
fails,  an adapted cavity approach
yields valid results. The learning curves, obtained in a suitable
thermodynamic limit, display a
first order phase transition from poor to perfect generalization.
\end{abstract}

\newcommand{\D}{{\mathbb D}}

During the last decade, independent component analysis (ICA) has emerged as
one of the most powerful unsupervised learning procedure for many
signal processing tasks \cite{Hyv01,Cic02}. It assumes that the observed, 
often high dimensional signal, is a linear mixture of {\em independent}
source signals and aims to recover these sources just from
observing the mixed up signal. Hence, ICA is sometimes also
called blind signal deconvolution. An illustrative scenario is the
cocktail party problem where, to understand any single speaker, we first
need to identify her voice amidst the jumble of sounds reaching our
ears. 

The basic finding in ICA is that the distribution of the observed
signal will be similar to a Gaussian, especially when
many independent sources contribute to the linear mixture. The source
signals, however, will often be highly structured, and
non-Gaussian. ICA thus searches for a linear transformation of the 
observations which maximizes non-Gaussianity by evaluating a suitable
contrast function. To detect this, the 
contrast function used must compute a higher than quadratic statistics of the 
transformed data.

In a principled way, ICA can be derived by considering the mutual
information of the transformed data, which is a natural measure of statistical 
dependence. To avoid the problem of density estimation, which 
arises in a direct evaluation of the mutual information, one then uses 
expansions (Edgeworth, Gram-Charlier) around Gaussianity to
approximate the mutual information \cite{Com94,Ama95}. 
This leads to  contrast
functions which are related to the higher order cumulants of
the transformed data.  

This Letter provides a first analysis of ICA for
polynomial  contrast functions using the
statistical physics of disordered systems.
Surprisingly,
the replica method, one of the most powerful tools in analyzing
quenched disorder, fails since it cannot  control the contributions to
the contrast function in the large deviations regime. However, a
physically valid analysis is obtained by adapting the cavity
method, showing that the scale of the learning curve depends on the
degree of  the polynomial. Unusually, for a system with continuous couplings,
the curve itself is a step function, jumping from poor to perfect 
generalization. But a badly generalizing state is always
metastable and it is remarkable that we can nevertheless find polynomial time
algorithms which generalize well.

In formal terms, we assume that the
observable signal $\xi$ can be written as $\xi = M\hxi$, where 
the source $\hxi$ is an $N$-dimensional  random variable with 
independent components and $M$ is the $N$ by $N$ mixing matrix.
Learning is based on a training set $\D$ of $P$ independent
observations  $\xi^\mu$
of the signal $\xi$, obtained for a fixed, if unknown, mixing matrix $M$.
 The deconvolution problem (finding $\hxi$)
can be decomposed by first finding just one independent component,
subtracting it from the mixture, and reapplying the procedure to the
remaining $N-1$ dimensional task. Hence, I shall just deal with
finding the first  component $\hxi_1$ and assume that it is non-Gaussian 
whereas all other components of $\hxi$ are Gaussian. 

Normally, the first step in ICA is to whiten the data, so that it has
zero mean and its covariance matrix is the identity. So, I shall
further assume that the source components have zero mean and unit
variance and that $M$ is orthogonal, $M^TM = \mathbf 1$. In short, the
ICA task now is to find, based on the training set $\D$, a vector $J$
such that $J^T\xi = \pm\hxi_1$. For this,
one picks a suitable non-quadratic contrast function $g$, computes the
empirical contrast
\begin{equation}
c_{\D}(J) = P^{-1} \sum_{\mu=1}^P g(J^T M\hxi^\mu), \label{contrast}
\end{equation}
and  chooses $J$ to maximize $c_{\D}(J)$ under the constraint $|J|=1$.
To analyze this problem,  one will
first  consider the Gibbs weight 
$\exp(\beta N c_{\D}(J))$ at some finite inverse temperature $\beta$
and calculate the typical value of the logarithm of its partition function 
$Z_\D =  \int {\rm d}J \exp(\beta N c_{\D}(J))$, where the integration
is over the uniform density on the unit sphere in $\R^N$. Since, via a
gauge, the  partition function is independent of the mixing matrix $M$,
we set $M= \mathbf 1$ for the analysis. 
 
I shall first consider the replica approach to this calculation and
for brevity assume that the contrast function is 
$g(x) = x^3$. We are then immediately faced with the problem that
the moments $\La Z_\D^n \Ra_\D$ do not exist, indeed $Z_\D$ does not
even have a mean 
\footnote{In a sense, this problem already crops up for principal
component analysis where $g(x)=x^2$. Then $\La Z_\D^n \Ra_\D$
diverges, if $n$ or $\beta$ are large enough. So, using replicas, one
is in effect computing a continuation from small $\beta$ and large $n$
to large $\beta$ and small $n$.
}.
A second issue arises since $c_{\D}(J)$ is ${\cal
O}(N^{3/2}/P)$ for $J = \xi^\mu/|\xi^\mu|$. So, if we have just 
$P = \alpha N$ examples, $\ln Z_\D$ is not an extensive quantity for
large $N$.

\newcommand{\KN}{K_{\!\scriptscriptstyle N}}
\newcommand{\LN}{L_{\scriptscriptstyle N}}
\newcommand{\gN}{g_{\scriptscriptstyle N}}

To address the first problem, we introduce a cutoff $\KN > 0$, replacing 
$g(x) = x^3$ by $\gN(x) = \max\{x^3,\KN^3\}$ in Eq. (\ref{contrast}). 
Since we want to
ultimately recover the $g(x) = x^3$ case, we assume that $\KN$
diverges with increasing $N$. 
Nevertheless, due to
the cutoff, the moments of  $Z_\D$ now exist for any finite $N$.
Further, we assume that the training set has $P=\alpha \LN N$ and
not just $\alpha N$ patterns. Then, if $\LN$ diverges sufficiently quickly
w.r.t. $N$ and $\KN$,  $\ln Z_\D$ will be an extensive quantity.
Finally, we should find that for the purpose of calculating  $\ln
Z_\D$ for large $N$, choosing $K_N = \sqrt{N}$ is equivalent to not
cutting off at all. The reason for this quite simply is that 
for $N\rightarrow\infty$
the fields $J^T \xi^\mu$ are bounded by $\sqrt{N}$ for
almost all training sets.  

In this setting, standard arguments yield the exact finite $N$ result 
\begin{eqnarray*}
\La Z_\D^n \Ra_\D &=& 
\lambda_{N,n}\!\! \int\!\! {\rm d}R{\rm d}Q 
  \det(Q\!-\! R R^T)^{\frac{N-n+1}{2}}
{\cal G}_{\scriptscriptstyle N} (R,Q)^N \\
{\cal G}_{\scriptscriptstyle N}(R,Q) &=& 
\La \prod_{a=1}^n
  \exp\left( \frac{\beta\max\{(R^a \xi_1 + X^a)^3,\KN^3\}}{\alpha L_N} 
\right)
  \Ra_{\xi_1,X}^{\alpha \LN}  
\end{eqnarray*}
Here $R$ is an $n$-vector, Q a symmetric $n$ by $n$ matrix with 
$Q^{aa}=1$, and the domain of integration is such that the matrix 
$Q - R R^T$ is positive definite.
The $X^a$ are zero mean Gaussian with covariances 
$\La X^a X^b\Ra = Q^{ab} - R^a R^b$, and $\lambda_{N,n}$ is obtained using that
the moments equal $1$  for $\beta = 0$.
Now, given any sequence of cutoffs
$\KN$, we can certainly find $\LN$ so that 
${\cal G}_{\scriptscriptstyle N}(R,Q)$ stays
finite for large $N$. Then, we should be able to use Laplace's method
of the maximum point to find that in the large $N$ limit
\begin{equation}
\frac{1}{N}\ln\La Z_\D^n \Ra_\D \!=\! \sup_{R,Q}\, 
\ln {\cal G}_N(R,Q) + \half \ln \det(Q\!-\! R R^T)\,. \label{lapl}
\end{equation}
But at this point, at the latest, it is clear that something is amiss.
The limiting value of the above RHS depends only on the
relative scalings of $K_N$ and $L_N$ and not on the relationship of
these scalings to the system size $N$. 
So (\ref{lapl}) implies  that the scale of learning curve can be
{\em arbitrarily} stretched by using cutoffs which diverge quickly
with $N$. This problem arises regardless of assumptions about replica
symmetry.

We proceed anyway and, using the replica symmetric
parameterization of (\ref{lapl}), find for $N\rightarrow\infty$
\begin{eqnarray}
\frac{1}{N}\La\ln Z_{\D} \Ra_\D
&=& 
\sup_r \inf_q\,\, G_r(q,R) + G_s(q,r) \nonumber \\
G_r(q,R) &=&
\alpha L_N \La \!\ln\!\La \exp\left(
    \frac{\beta}{\alpha L_N}\gN(r \xi_1 + \sqrt{q-r^2}y_0+\sqrt{1-q}y_1) 
       \right)  \Ra_{\!\!y_1} \Ra_{\!\!\xi_1,y_0} \nonumber \\  
G_s(q,r) &=& \half \frac{q-r^2}{1-q} + \half\ln(1-q) \label{rsZ} 
\end{eqnarray}
where 
$y_o,y_1$ are standard Gaussians, i.e. with zero mean, unit variance.
The extremal $r$ 
is just the typical value of the first component of a weight vector
picked from the Gibbs density and
measures to which extent the structure in the data is recognized.
Using (\ref{rsZ}), we relate the scalings of
$\KN$ and $\LN$. For $\LN \gg \KN$ the energy term converges to 
$G_r(q,R) = r^3 \La \xi_1^3 \Ra$. This is the limit of many
examples where $r=1$ for all $\alpha$. In contrast, for $\LN \ll \KN$
there are too few examples and  $G_r(q,R)$ diverges.

So, the scale of the learning curve is given by setting $\LN = \KN$.
On this scale, 
we find that  $G_r(q,R)$ converges to $r^3 \La \xi_1^3 \Ra$ as in the
limit of many examples if $q$ exceeds a critical value
$q_c(\alpha,\beta)$,  whereas $G_r(q,R)$ diverges for $q
<q_c(\alpha,\beta)$. Solving the extremal problem for $q$ by taking the
limit $q\rightarrow q_c(\alpha,\beta)$ from above, then taking the 
$\beta\rightarrow\infty$ limit, we finally find the
simple result for the
ground state:
$
c(\alpha)= \sup_r
r^3 \La  \xi_1^3 \Ra_{\xi_1}+  (1-r^2)/\alpha. 
$
Here $c(\alpha)$ is the typical value of the highest achievable
empirical contrast, $\max_{|J|=1} c_\D(J)$. The learning curve for $r$
thus obtained, is a step function showing a first order
phase transition at $\alpha_c = 1/\La  \xi_1^3 \Ra_{\xi_1}$ 
from no learning ($r=0$) to perfect learning ($r=1$).
But the $r=0$ state is metastable for all values $\alpha >
\alpha_c$.

\begin{figure}
   \begin{tabular}{l}
        \mbox{\begin{tabular}{l}
           \includegraphics[scale=0.8]{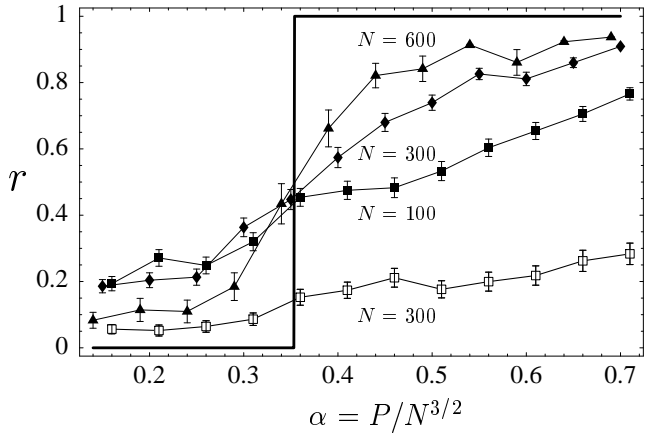}
              \end{tabular}}
   \end{tabular}   
 \caption{
 Prediction of $\KN=\sqrt{N}$ replica theory (bold line) compared to
 simulation results. The non Gaussian source is 
  $\hat\xi_1 =(y^2-1)/\sqrt 2$, where $y$ is a standard Gaussian.
 The empty symbols show the results for the algorithm finding local
 maxima of the empirical contrast. The full symbols, denoting results
 for the iterated version of the procedure described in the main text,
 show that the agreement with the replica theory improves quickly with
 increasing system size $N$ for this algorithm.
 The  error bars estimate the standard deviation of the sample to sample 
 fluctuations.
}
\end{figure}

The replica theory predicts that for any divergent sequence of
cutoffs $\KN$, e.g. $\KN = e^N$,  we need $P > \alpha_c \KN N$ examples for
good generalization when $N$ is large. 
While this is ridiculous, I have argued above
that choosing $\KN=\sqrt N$ is, for $N\rightarrow\infty$, 
 equivalent to not cutting off at all. To
compare the replica result for this choice of $\KN$ to
numerical  simulations, let us consider 
actually finding a weight vector maximizing $c_\D(J)$. 
It turns out that a rather simple discrete dynamics can be used since
$g(x) = x^3$. Starting with a random
vector of unit length $J^0$, at the $k$-th time step we first compute the
matrix 
$A(J^k) = \sum_{\mu=1}^P \xi^\mu ({J^k}^T \xi^\mu ) {\xi^\mu}^T$
and then choose $J^{k+1}$ to maximize 
$|J^T A(J^k) J|$ under the constraint $|J|=1$. 
So, $J^{k+1}$ is an
eigenvector to the eigenvalue of largest magnitude of $
A(J^k)$. Standard results on quadratic forms imply that 
$|{J^{k+1}}^T A(J^k) J^{k+1}| \geq   |{J^{k}}^T A(J^{k-1}) J^{k}|$,
and the inequality is strict unless we are at a fixed point. 
Hence, the iteration converges to a vector $J^\infty$ which is a local 
maximum or minimum of  $c_\D(J)$. In the latter case, we just flip the
sign of $J^\infty$ to obtain a local maximum. 

Simulation results for the procedure, compared to the $\KN =
\sqrt{N}$ replica theory in Fig. 1, show that the performance of
the algorithm is rather poor. This is in line with the
theoretical findings, since these predict that $r=0$ is
metastable, and the algorithm is only finding a local maximum. Figure 1
also shows result for an iterated version of the algorithm. There the
algorithm is rerun with $m=0.1N$ different random initial conditions,
and the weight vector maximizing $c_\D(J)$ among the $m$ outcomes is
chosen. These result are in good agreement with the $\KN =
\sqrt{N}$ replica theory, indicating that beyond the phase transition the
basin of attraction of the global maximum is quite large. 

Even if the simulations indicate
that the replica approach is saved by
in the end plugging in the correct scaling of the cutoff $\KN$,
the theoretical situation is highly unsatisfactory. 
I shall next show that a physically
reasonable analysis can be provided by adapting the cavity method.
This is much simplified if make some major
changes to the notation. From now on the non-Gaussian source will be
denoted by $\gamma$, whereas all of the $N$  components of $\xi$ are
assumed independent standard Gaussian. Our primary goal is to calculate
the typical value of $C_r = \max_{|J|=1} C_r(J)$ with
\begin{equation}
C_r(J) = \frac{1}{P}\sum_{\mu=1}^P g(r \gamma^\mu + \sqrt{1-r^2} J^T
\xi^\mu)
\label{orig}
\end{equation}
where $J$ is an N-dimensional vector. So $C_r$ is the maximal value of
the empirical contrast achievable on an $r$-shell. For generality, we
shall now longer assume that $g(x)$ must be cubic but consider any
super-quadratic function which does not diverge too quickly. 
In particular, for some $k>0$, 
$
\lim_{x\rightarrow\infty}{g(x)}/{x^{2+k}} = \psi 
$
should exist and be positive. Without loss of generality, we may then
assume $\psi=1$.

We still have $P=\alpha \LN N$
examples and consider the random variable $J_\D$ with the Gibbs
density    
\begin{eqnarray}
p_\D(J) &=& \frac{1}{Z_\D(\beta)} 
          \frac{e^{-\half |J|^2}}{(2 \pi)^{\half N}} 
          \prod_{\mu=1}^P 
            e^{\frac{\beta}{\LN} 
                     g(\gamma^\mu,[J]^T\xi^\mu)}
\nonumber \\
g(\gamma^\mu,[J]^T\xi^\mu) &=& 
g(r \gamma^\mu + \sqrt{1-r^2} [J]^T\xi^\mu)\,. \label{GD} 
\end{eqnarray}
Here $[J] = J/|J|$ and $Z_\D(\beta)$ is given
by the normalization $\int \!{\rm d}J\, p_\D(J) =1$. Note, that we are now
using a factorizing Gaussian prior on $J$ and, to compensate for this, the
normalized vector $[J]$ is used to calculate the field in (\ref{GD}). 

A key task in the cavity approach is  obtain the field distribution by
calculating  the thermal average 
$\La \phi(\gamma^\mu,[J_D]^T\xi^\mu) \Ra_{J_\D}$ for  any function
$\phi$. One finds 
\begin{eqnarray}
\La\phi(\gamma^\mu,[J_D]^T\xi^\mu) \Ra_{J_\D} &=& 
\frac{Z_{\D/\mu}(\beta)}{Z_\D(\beta)}
\La e^{\frac{\beta}{\LN} 
                     g(\gamma^\mu,[J_{\D/\mu}]^T\xi^\mu)}
    \phi(\gamma^\mu,[J_{\D/\mu}]^T\xi^\mu) \Ra_{J_{\D/\mu}},
\label{cav}
\end{eqnarray}
where $J_{\D/\mu}$ is the random variable with the Gibbs density obtained 
when pattern $\mu$ is removed from the system, i.e. 
omitting the $\mu$-th factor
of the product in (\ref{GD}) and adjusting the partition function to 
$Z_{\D/\mu}(\beta)$.
The variance of the cavity field 
$[J_{\D/\mu}]^T\xi^\mu$ is a self averaging quantity and it must then 
equal $1-q$ for large $N$, where 
$q = |\La [J_{\D/\mu}] \Ra_{J_{\D/\mu}}|^2$. Normally, one would further argue
that  $[J_{\D/\mu}]^T\xi^\mu$ becomes Gaussian in the thermodynamic limit.
But if we assume this, 
the $J_{\D/\mu}$ average in (\ref{cav}) diverges even when 
$\phi$ is a simple bounded function. 
This highlights the fact that the cavity field is not Gaussian in the large 
deviations regime because
$[J_{\D/\mu}]^T\xi^\mu$ cannot be larger than $|\xi^\mu|$.

Hence, I rephrase the cavity argument as follows: For the purpose of 
calculating overlaps with a random vector such as $\xi^\mu$, 
the not normalized $J_{\D/\mu}$ can for large $N$ be treated as a
Gaussian (with covariance matrix $(1-q)\mathbf 1$).
Then, the fluctuations of the cavity field obtained using
the normalized $[J_{\D/\mu}]$,
\[
P_{N,q}(h) = \La 
\delta\left(h - 
\left([J_{\D/\mu}]^T-\La[ J_{\D/\mu}]^T\Ra_{J_{\D/\mu}}\right)\xi^\mu
\right) \Ra_{J_{\D/\mu}}
\]
can be explicitly calculated.
This yields the 
important fact that there are just two relevant scales for the cavity 
fluctuations. 
For large $N$, 
$P_{N,q}(h)$ converges to 
$e^{-\half h^2/(1-q)}/\sqrt{2 \pi (1-q)}$ 
if  $h \ll \sqrt{N}$,  but in the large deviations regime, for
$h = d \sqrt{N}$,
\begin{equation}
\lim_{N\rightarrow\infty} N^{-1}\ln P_N(d \sqrt{N}) =
-\half \frac{ q d^2}{1-q} + \half\ln(1-d^2)
\label{ldev}
\end{equation}
if $0\leq d\leq1$.
Now, in terms of the functional
\[
{\cal L}^{q,\beta}_{y,\gamma}(\phi) = 
\int_{-\sqrt{N}}^{\sqrt{N}} 
{\rm d}h\, P_{N,q}(h)\,\phi(\gamma,\sqrt{q}y+h)\,
e^{\frac{\beta}{\LN} g(\gamma,\sqrt{q}y+h)}
\]
the average in Eq. (\ref{cav}) can in the limit of large $N$  be rewritten as 
$\La\phi(\gamma^\mu,[J_D]^T\xi^\mu) \Ra_{J_\D} =
{\cal L}^{q,\beta}_{y^\mu,\gamma^\mu}(\phi)/
{\cal L}^{q,\beta}_{y^\mu,\gamma^\mu}(1)$ 
with $y^\mu = q^{-\half}\La[ J_{\D/\mu}]\Ra_{J_{\D/\mu}}^T\xi^\mu$. So  the 
quenched  averages are
\begin{eqnarray}
\La \La \phi(\gamma^\mu,[J_D]^T\xi^\mu) \Ra_{J_\D} \Ra_\D
&=& \La
\frac{{\cal L}^{q,\beta}_{y,\gamma}(\phi)}
     {{\cal L}^{q,\beta}_{y,\gamma}(1)} \Ra_{y,\gamma}  \label{qav} \\
\La \ln Z_\D(\beta) - \ln Z_{\D/\mu}(\beta) \Ra_\D &=&
\La \ln {\cal L}^{q,\beta}_{y,\gamma}(1)  \Ra_{y,\gamma} 
\label{qav1}
\end{eqnarray}
where $y$ is standard Gaussian. The last equation is 
obtained by setting $\phi =1$ in (\ref{cav}).

We can now consider whether the large deviations regime contributes to
the averages in (\ref{qav}) for a polynomially bounded
$\phi$. Using that for large arguments $g(x) \sim x^{2+k}$ and
referring to  Eq. (\ref{ldev}), we find that it
will contribute if the maximum of 
\begin{equation}
u(d) = 
\beta d^{k+2}\frac{N^{\half k}}{\LN} 
- \half \frac{ q d^2}{1-q} + 
\half\ln(1-d^2)
\label{reldev}
\end{equation}
is positive for large $N$. This won't happen if
$\LN \gg  N^{\half k}$ and 
Eq. (\ref{qav}) then implies that
$\La \La \phi(\gamma^\mu,[J_D]^T\xi^\mu) \Ra_{J_\D} \Ra_\D =
\La \phi(\gamma,y) \Ra_{y,\gamma}$. The empirical mean equals
the expectation value and so the learning curve is  trivial. 
Henceforth, we focus on the relevant scale, setting 
$\LN =  N^{\half k}$. 

Our next task is to calculate the response when a new coupling $J_0$ is 
added to the system and each pattern $\xi^\mu$ is augmented by
a new component $\xi_0^\mu$. We denote the augmented training set by
$\hat\D$ and use (\ref{GD}) to define the partition function  
$Z_{\hat\D}(\beta)$ of the $N+1$ dimensional system. 
Due  to the $N$-dependence of the Gibbs weight 
$e^{\frac{\beta}{\LN}g(\gamma^\mu,[J]^T\xi^\mu)}$, it is simplest
to assume a slightly different temperature 
$\hat\beta_N = \beta L_{\scriptscriptstyle N+1}/\LN$ 
in the augmented system. Then,
when  considering the ratio $Z_{\hat\D}(\hat\beta_N)/Z_{\D}(\beta)$,
the two systems have the same Gibbs weight per pattern.
Standard arguments  \cite{Mez89} thus apply and yield that 
$
\sLa \ln Z_{\hat\D}(\hat\beta_N)/Z_\D(\beta) \sRa_{\hat\D}
= G_s(q,0)\, \label{entres}
$ for large $N$.
Here $G_s(q,0)$ is the entropy term of the
replica theory (Eq. \ref{rsZ}), but evaluated at $r=0$ because we are
calculating the partition function for each $r$-shell individually.

Having identified, via $\LN=
N^{\half k}$, the scale of the learning curve, 
$N^{-1}\La \ln Z_\D(\beta) \Ra_D$ will 
converge to a finite quantity  $z(\alpha,\beta)$ in the thermodynamic limit.
We then  have
\newcommand{\pdev}[1]{ \frac{\partial\,\,}{\partial #1} } 

\begin{eqnarray*}
  \sLa \ln Z_{\hat\D}(\hat\beta_N)/Z_\D(\beta) \sRa_{\hat\D} &=&
 z(\alpha,\beta) - 
 \alpha \frac{k+2}{2}\pdev{\alpha}{z(\alpha,\beta)} +
 \frac{\beta k}{2} \pdev{\beta}{z(\alpha,\beta)}.                   
\end{eqnarray*}
The derivative of $z$ with respect to $\alpha$ is obtained from
Eq. (\ref{qav1}), and the thermal derivative is found
from  (\ref{qav}) using $\phi =g$. 

Putting things together, we finally find for large $N$
\begin{eqnarray}
z(\alpha,\beta) &=&   
\La \alpha\frac{k+2}{2} N^{\half k}  \ln {\cal L}^{q,\beta}_{y,\gamma}(1)
- \frac{\beta k}{2} \frac{{\cal L}^{q,\beta}_{y,\gamma}(g)}
                          {{\cal L}^{q,\beta}_{y,\gamma}(1)}
\Ra_{y,\gamma}\!\! 
+ G_s(q,0)\,, \label{zfunc}
\end{eqnarray}
where the value of $q$ still has to be determined.

For this, let us reconsider when the large deviations regime
contributes to the value of ${\cal L}^{q,\beta}_{y,\gamma}(1)$. Going back
to Eq. (\ref{reldev}), with $\LN =  N^{\half k}$, 
we see that as in the replica theory this is governed by a critical
value $q_{\rm c}(\beta)$ of $q$. 
For $q < q_{\rm c}(\beta)$, $\max_d u(d)$ is positive in the large $N$ limit, 
so (\ref{zfunc}) diverges.
The possible range for $q$ is thus $q_{\rm c}(\beta) \leq q \leq 1$.
But,  if we assume $q > q_{\rm c}(\beta)$, the large $N$ limit yields the
very simple result
$
z(\alpha,\beta) =   G_s(q) + \alpha \beta \La g(\gamma,y) \Ra_{\gamma,y}
$. 
Now, on one hand,  the empirical contrast is found by
differentiating $z(\alpha,\beta)$ w.r.t to $\beta$. This yields  
$\La g(\gamma,y) \Ra_{\gamma,y} + \frac{1}{\alpha}G'_s(q)\pdev{\beta}q$.
But computing the same quantity using (\ref{qav}) yields
$\La g(\gamma,y) \Ra_{\gamma,y}$. So $q$ must stay  constant when $\beta$ 
varies, but this is impossible since $q_{\rm c}(\beta)\rightarrow 1$ for 
$\beta\rightarrow\infty$.

Hence, the only possible value for $q$ is $q_{\rm
c}(\beta)$.
Evaluating (\ref{zfunc}) by taking the limit $q\rightarrow q_{\rm
c}(\beta)$ from above, leads to the same result as in the $\KN =
\sqrt{N}$ replica theory. But, of course, this  has the same 
inconsistencies as found for the $q > q_{\rm c}(\beta)$ assumption.
It also makes no physical sense to use (\ref{zfunc})
at the point of discontinuity since  the cavity  derivation neglects 
fluctuations of $q$. Even if these vanish with increasing $N$, at the point
of discontinuity, $q=q_{\rm c}(\beta)$, the true result will 
nevertheless  depend on the unknown fluctuations.

But some conclusions can be drawn, knowing that $q$ has the
critical value. Let $d_\beta$ be the unique positive value such that 
$u(d_\beta) =0$ for   $q=q_{\rm c}(\beta)$. Then arguments analogous
to the derivation of  (\ref{qav}) show that the probability of the
posterior field $[J_\D]\xi^\mu$ exceeding $d\sqrt{N}$ is {\em not}
exponentially small if $d$ is lower than $d_\beta$. 
More precisely, one finds for
$N\rightarrow\infty$ and $d < d_\beta$
\begin{eqnarray*}
\La N^{-1}\ln\sLa \Theta([J_D]^T\xi^\mu - d\sqrt{N}) \sRa_{J_\D} \Ra_\D
&=&  \\ 
\La N^{-1}\ln 
{{\cal L}^{q,\beta}_{y,\gamma}(\Theta(h - d\sqrt{N}))}/
     {{\cal L}^{q,\beta}_{y,\gamma}(1)} \Ra_{y,\gamma} &=& 0\,.
\end{eqnarray*}
Further, $d_\beta$ approaches $1$ with increasing $\beta$. But this is
only possible if simply aligning the weight vector with the pattern $\xi^\mu$
maximizes the empirical contrast, at least upto sub-extensive corrections. So,
in the notation of Eq. \ref{orig}, we have $C_r = C_r([\xi^\mu ])$
for large $N$, and thus finally
\begin{equation}
C_r  =
(1-r^2)^{\frac{2+k}{2}}/\alpha + \La g(r \gamma + \sqrt{1-r^2}\,y) 
\Ra_{\gamma,y}\,. 
\label{final}
\end{equation}
Maximizing this in $r$, the same learning curve is obtained for
the cubic case, $g(x)=x^3$, as in
the  $\KN=\sqrt N$ replica theory
\footnote{
For $g(x)=x^4$, the curve depends on whether $\sLa \gamma^4
\sRa_\gamma > 3$, since the fourth moment of a standard Gaussian is
$3$. If so, the value of $r$ jumps from $0$ to $1$ at 
$\alpha_c = 1/(\sLa \gamma^4\sRa_\gamma - 3)$. The 
$\sLa \gamma^4\sRa_\gamma < 3$ case, where one will use  $g(x)=-x^4$,
shall be described elsewhere. It 
is much simpler since the large deviations regime does not contribute.}.
It is important to note that we have in essence just used the standard
weak correlation assumptions of the cavity method in deriving (\ref{final}).
In view of the good agreement with numerical simulations (Fig. 1),
this strongly suggests that the cavity result is indeed exact in the 
thermodynamic limit.

From an analytical point of view, it is intriguing that the present
problem reveals a difference in the scope of the replica and the
cavity method. The latter can be transparently adapted to take
into account that the cavity field is not Gaussian in the large
deviations regime. But, commuting the thermal average with the disorder
average, at the expense of considering moments, is part and parcel
of using replicas. As a consequence, all the relevant fields
become truly Gaussian. This points to implicit assumptions in the
replica method, which need to be taken care of in any program to put
the approach on a solid mathematical footing \cite{Par02}. 

\acknowledgements  
 
It is a pleasure to acknowledge many discussions with Manfred Opper.
This work was supported by the Deutsche Forschungsgemeinschaft.

\bibliographystyle{unsrt}
\bibliography{/home/robert/tex/neural}

\begin{thebibliography}{1}

\bibitem{Hyv01}
A.~Hyv{\"a}rinen, J.~Karhunen, and E.~Oja.
\newblock {\em Independent Component Analysis}.
\newblock Wiley-Interscience, 2001.

\bibitem{Cic02}
A.~Cichoki and S.~Amari.
\newblock {\em Adaptive Blind Signal and Image Processing}.
\newblock John Wiley \& Sons, 2002.

\bibitem{Com94}
P.~Comon.
\newblock Independent component analysis, a new concept?
\newblock {\em Signal Processing}, 36:287 -- 314, 1994.

\bibitem{Ama95}
A.~Amari, A.~Cichoki, and H.H. Yang.
\newblock A new learning algorithm for blind source separation.
\newblock In {\em NIPS 8}, pages 757--763. MIT Press, 1996.

\bibitem{Mez89}
M.~M{\'e}zard.
\newblock The space of interactions in neural networks: Gardner's computation
  with the cavity method.
\newblock {\em J. Phys. A}, 22:2181--2190, 1989.

\bibitem{Par02}
G.~Parisi.
\newblock {\em http://arxiv.org/abs/cond-mat/0207334}, 2002.

\end{thebibliography}

\end{document}